\documentclass[pre,aps,amsfonts,amssymb,twocolumn,epsfig,graphics,referee]{revtex4}

\def \beq{\begin{equation}}         \def \eeq{\end{equation}}
\def \beqa{\begin{eqnarray}}        \def \eeqa{\end{eqnarray}}
\def \bea{\begin{array}}        \def \eea{\end{array}}

\def\bio#1#2#3{{\it Biophys. J. }{\bf #1}, #2, #3}
\def\cell#1#2#3{{\it Cell} {\bf #1}, #2, #3}

\def\jcp#1#2#3{{\it J. Chem. Phys. }{\bf #1}, #2, #3}
\def\nat#1#2#3{{\it Nature }{\bf #1}, #2, #3}

\def\pnas#1#2#3{{\it Proc. Natl. Acad. Sci. USA }{\bf #1}, #2, #3}
\def\pre#1#2#3{{\it Phys. Rev. E }{\bf #1}, #2, #3}

\def\sci#1#2#3{{\it Science }{\bf #1}, #2, #3}

\usepackage{amsmath}
\usepackage{epsfig}
\begin{document}

\title{Bell Rate Model with Dynamic Disorder: Model and Its Application
in the Receptor-ligand Forced Dissociation Experiments}
\author{Fei Liu$^{1\ast}$ and Zhong-can Ou-Yang$^{1,2}$}
\address{$^{1}$Center for Advanced Study, Tsinghua University, Beijing, China}
\address{$^{2}$Institute of Theoretical Physics,
The Chinese Academy of Sciences, P.O.Box 2735 Beijing 100080,
China}
\email[Email address:]{liufei@tsinghua.edu.cn}
\date{\today}

\begin{abstract}
We extend the Bell forced dissociation rate model to take account
into dynamic disorder. The motivation of the present work is from
the recent forced dissociation experiments of the adhesive
receptor-ligand complexes, in which some complexes were found to
increase their mean lifetimes (catch bonds) when they are
stretched by mechanical force, while the force increases beyond
some thresholds their lifetimes decrease (slip bonds). Different
from our previous model of force modulating dynamic disorder, in
present work we allow that the projection of force onto the
direction from the bound to the transition state of complex could
be negative. Our quantitative description is based on a
one-dimension diffusion-assisted reaction model. We find that,
although the model can well describe the catch-slip transitions
observed in the single bond P-selctin glycoprotein ligand
1(PSGL-1)$-$P- and L-selectin forced dissociation experiments, it
might be physically unacceptable because the model predicts a
slip-catch bond transitions when the conformational diffusion
coefficient tends to zero.

 \end{abstract}

\maketitle

\section{Introduction}
Traditionally, the forced dissociation rate of noncovalent
biological receptor-ligand bonds is described by the simple Bell
expression~\cite{Bell}
\begin{eqnarray}
k_{\rm off}(f)=k^0_{\rm off}\exp(f\xi^{\ddag}/k_{\rm B}T),
\label{Bellmodel}
\end{eqnarray}
where
\begin{eqnarray}
k^0_{\rm off}=k_0\exp(-\Delta G^\ddag/k_{\rm B}T)
\end{eqnarray}
is the intrinsic rate constant in the absence of force, $\Delta
G^\ddag$ is the height of the intrinsic energy barrier,
$\xi^{\ddag}$ is a projection of the distance from the bound state
to the energy barrier onto external applied force $f$, $k_{\rm B}$
is the Boltzmann's constant, and $T$ is absolute temperature. Bell
phenomenologically introduced this expression from the kinetic
theory of the strength of solids in his seminal paper more than
twenty years ago. Evans and Ritchie~\cite{Evans97} later put the
Bell expression onto a firm theoretical bases: the bond rupture
was modelled in the framework of Kramers theory as thermally
assisted escape over one or several transition state barriers. The
validity of the expression has been demonstrated in
experiments~\cite{Alon,Chen}. In the Bell forced dissociation rate
model, the parameters including the energy barrier $\Delta
G^\ddag$ and the projection distance $\xi^\ddag$ are deterministic
and time independent.

On the other hand, we have known that in a large variety of
chemical and physical areas, association/dissociation processes
could be stochastic and time dependent. Such processes have been
termed as ``rate processes with dynamical disorder" in an
excellent review given by Zwanzig~\cite{Zwanzig} in fifteen years
ago. The typical examples include cyclization of polymer chains in
solutions~\cite{Wilemski}, ligands rebinding to heme
proteins~\cite{Agmon}, electron transfer reactions~\cite{Sumi},
electronic relaxation in solutions in the absence of an activation
barrier~\cite{Bagchi}, and etc. Hence, in principle we could
expect that dynamic disorder may also arise in some forced
biological bond rupture experiments. But it was not until recently
that we~\cite{liuf1,liuf2} firstly pointed that this disorder
might play key role in a forced dissociation experiment of the
adhesive bond forming between P-selectin and P-selectin
glycoprotein ligand 1 (PSGL-1) using atomic force microscopy
(AFM)~\cite{Marshall}, though this experiment was published three
years ago. A counterintuitive observation has been found: in
contrast to ordinary biological bonds, the mean lifetimes of the
PSGL-1$-$P-selectin bond first increase with initial application
of a small force, which were termed ``catch" by Dembo~\cite{Dembo}
early, and subsequently decrease, which were termed ``slip".

The main point of the questions of dynamic disorder is that the
rate coefficient depends on stochastic control variables and is
fluctuating in time~\cite{Zwanzig}. Correspondingly, we suggested
that the Bell parameters in Eq.~\ref{Bellmodel} were
stochastic~\cite{liuf1,liuf2}. As one type of noncovalent bonds,
interactions of adhesive receptor and their ligand is always
weaker. Moreover, the interface between them has been reported to
be broad and shallow, such that revealed by the crystal structure
of the PSGL-1$-$P-selectin bond~\cite{Somers}. Therefore it is
plausible that the energy barriers $\Delta G^\ddag$ or the
projection distance $\xi^\ddag$ for the bond are fluctuating with
time due to either the global conformational change or the local
conformational change at the interface. Two possible cases were
discussed. In the Gaussian stochastic rate model
(GSRM)~\cite{liuf1}, we simply assumed that $\xi^\ddag$ and
$\Delta G^\ddag$ had basic properties of random variables. We
concluded that the catch behavior of some biological adhesive
bonds could arise from the stronger positive correlation between
the two stochastic variables. Although the model accounted for the
catch-slip bond transition in a direct and analytic way, it also
predicted that the mean lifetime of the PSGL-1$-$P-selectin bond
was symmetric relative to a critical force $f_c$, while the
experimental data were clearly skewed towards a large force. In
addition, the physical justice of the assumption of a stationary
dissociation process may be doubtful. To overcome these shortages,
we proposed an alternative and more ``microscopic"
model~\cite{liuf2}, in which external applied force not only
presents in the Bell expression, but also modulates the
distribution of a ``inner" conformational coordinate. We named it
as force modulating dynamic disorder (FMDD) model in the
following. In contrast to the GSRM, we only allowed the energy
barrier to be stochastic. The quantitative description was based
on a one-dimension diffusion-reaction equation~\cite{Agmon}. We
found that agreement between our calculation and the data was
impressive. This model also suggests a new physical explanation of
the catch-slip bond: the transition could arise from a competition
of the two components of applied force along the directions of the
dissociation reaction coordinate and the complex conformational
coordinate; the former accelerates the dissociation by lowering
the height of the energy barrier, i.e., the slip behavior, while
the later stabilizes the complex by dragging the system to the
higher barrier height, i.e., the catch behavior.

Even there are apparent advantages of the FMDD model, we cannot
completely exclude the GSRM, because compared to the former, the
GSRM does not require that force acts on the ``inner" conformation
coordinate. This requirement makes us to conclude that catch-slip
bond transitions would be altered by the orientation of external
forces~\cite{liuf2}. Although such a prediction has a potential to
account for the bulk experimental observations~\cite{McEver} that
the optimal binding of P-selectin is critically dependent on the
relative orientations of its ligand, there is no such evidence in
the existing data measured by the single molecular
techniques~\cite{Marshall,Evans2004}. On the other hand, we have
viewed the conformational coordinate to be molecular extension
which is coupled to external applied force. We also have no strong
experimental evidence to support this assumption. The inner
coordinate has been thought to be experimentally unaccessible. In
addition, the FMDD model did not investigate the possibility of a
fluctuating projection distance. In the present work, we try to
give a model that force does not modulate the conformational
coordinate of the complex while presenting catch-slip behaviors
simultaneously. Our model still bases on a diffusion-reaction
equation. Different from the FMDD model, the projective distance
$\xi^\ddag$ is allowed to be fluctuating and negative. Our
calculation shows that this new Bell rate model with dynamic
disorder not only derives the same expression of the mean lifetime
which has been obtained in the GSRM before if both $\Delta
G^\ddag$ and $\xi^\ddag$ are simple linear functions of the
conformational coordinate, but also well fits to the experimental
data if a slightly complicated piecewise function with two
segments is used. An unexpected consequence of this model is that
the transition from catch to slip converts into slip-catch
transition when the conformational diffusion is ``frozen" by
lowering the diffusion coefficient to zero. Interestingly, the
FMDD model does not predict such a conversion. Therefore, our
discussion here would be useful to design new experiments to
distinguish which model is physically correct.

The organization of the paper is as follows. In Sec II, we
describe the physical picture of the Bell rate model with dynamic
disorder and give the essential mathematic derivations. According
to the value of the diffusion coefficient, we distinguish three
cases: $D\to\infty$, $D\to 0$, and the intermediate $D$.
Analytical solutions can be obtained for the first two cases,
while for the last one we resort to numerical approach. In the
following section, we study two extended Bell models at the three
different coefficients, in which the barrier height and projection
distance are linear and piecewise functions of the conformational
coordinate, respectively. A comparison between model and the
experiments is also performed. Finally we give our conclusion in
Sec IV.

\section{Bell rate model with dynamic disorder}
The physical picture of our model for the forced dissociation of
the receptor-ligand bonds is very similar with the small ligand
binding to heme proteins~\cite{Agmon}: there is a energy surface
for the dissociation which dependents on both the dissociation
reaction coordinate of the receptor-ligand bond and the
conformational coordinate $x$ of complex, while the later is
perpendicular to the former; for each complex conformation $x$
there is a different dissociation rate constant which obeys the
Bell expression $k_{\rm off}(x,f)$. Higher temperature or larger
diffusivity (low viscosities) allows $x$ variation within the
complex to take place. Conversely, at very low temperatures (or
high viscosities) the coordinate is frozen so that the complex is
ruptured with the rate constant $k_{\rm off}(x,f)$ without
changing their $x$ value during dissociation.

\subsection{Rapid diffusion}
{\bf The constant force mode.} There are two types of force
loading modes. First we consider the constant force
case~\cite{Marshall,Sarangapani}. A diffusion equation in the
presence of a coordinate dependent reaction is given
by~\cite{Agmon}
\begin{eqnarray}
\label{origindiffusionequation} \frac{\partial p(x,t)}{\partial
t}=D\frac{\partial^2 p}{\partial x^2}+ \frac{D}{k_{\rm B}T}
\frac{\partial}{\partial x}\left(p\frac{\partial V}{\partial
x}\right)-k_{\rm off}(x,f)p,
\end{eqnarray}
where $p(x,t)$ is the probability density for finding a value $x$
at time $t$, at initial time $p(x,0)$ is thought to be thermal
equilibrium under a potential $V(x)$ without reaction, and $D$ is
a constant diffusion coefficient. The motion is under influence of
a potential $V(x)$ and a coordinate dependent Bell expression
$k_{\rm off}(x,f)$. Because the above equation in fact is almost
same with that proposed by Agmon and Hopfield except that the rate
constant is controlled by applied force $f$, we only present
essential mathematic derivations. More detailed discussion about
the diffusion-reaction equation could be found in their original
work.

Following Agmon and Hopfield~\cite{Agmon}, we substitute
\begin{eqnarray}
\label{convertingfunction}
p(x,t)=N_0\phi(x,t)\exp\left[-V(x)/2k_{\rm B}T\right]
\label{transform}
\end{eqnarray}
into Eq.~\ref{origindiffusionequation}, one can convert the
Eq.~\ref{origindiffusionequation} into the Schr$\ddot
o$dinger-like presentation,
\begin{eqnarray}
\frac{\partial \phi}{\partial t}=D\frac{\partial^2 \phi}{\partial
x^2}-U_f(x)\phi=-{\cal H}_f(\phi),\label{Schodingerequation}
\end{eqnarray}
where $N_0$ is the normalization constant of the density function
at $t=0$, and the ``effective" potential
\begin{eqnarray}
\label{forcedependentquantumpotential}
U_f(x)&=&U(x)+k_{\rm off}(x,f)\\
&=&\frac{D}{2k_{\rm B}T}\left[\frac{1}{2k_{\rm
B}T}\left(\frac{dV}{dx}\right)^2-\frac{d^2 V}{d x^2}\right]+k_{\rm
off}(x,f).\nonumber
\end{eqnarray}
We define $U(x)$ for it is independent of the force. In principle
Eq.~\ref{Schodingerequation} could be solved by eigenvalue
technique~\cite{Agmon}. Particularly, at larger $D$ only the
smallest eigenvalue $\lambda_0(f)$ mainly contributes to the
eigenvalue expansion. For any given $k_{\rm off}(x,f)$, such as
the exponential form in the Bell expression, there is no
analytical $\lambda_0(f)$; instead a perturbation approach has to
be applied~\cite{Messiah}. If the eigenfunctions and eigenvalues
of the ``unperturbed" Schr$\ddot o$dinger operator,
\begin{eqnarray}
{\cal H}=-D\frac{\partial^2}{\partial x^2}+U(x),
\end{eqnarray}
in the absence of $k_{\rm off}(x,f)$ have been known to be
\begin{eqnarray}
{\cal H}\phi^0_n=-\lambda^0_n\phi^0_n,
\end{eqnarray}
and the Bell rate is adequately small (compared to the diffusion
coefficient $D$), the first eigenfunction $\phi_0(f)$ and
eigenvalue $\lambda_0(f)$ of the operator ${\cal H}_f$ then are
given by
\begin{eqnarray}
\label{eigenfunctionexpansion}
\phi_0(f)&=&\phi^{(0)}_0+\phi^{(1)}_0(f)+\cdots\\
&=&\phi^0_0+\sum_{m\neq 0}\frac{\int\phi^0_0(x)k_{\rm
off}(x,f)\phi^0_m(x)dx}{\lambda^0_0-\lambda^0_m}\phi^0_m+\cdots\nonumber
\end{eqnarray}
and
\begin{eqnarray}
\label{eigenvalueexpansion}
\lambda_0(f)&=&\lambda^{(0)}_0+\lambda^{(1)}_0(f)+\lambda^{(2)}_0(f)+\cdots\\
&=&\lambda^0_0+\int\phi^0_0(x)k_{\rm off}(x,f)\phi^0_0(x)dx+\nonumber\\
&&\sum_{m\neq 0}\frac{\left(\int \phi^0_0(x)k_{\rm off}(x,f)
\phi^m_0(x)dx\right)^2}{\lambda^0_0-\lambda^0_m}+\cdots,\nonumber
\end{eqnarray}
respectively. Because the system is assumed to be thermal
equilibrium at $t=0$, the first eigenvalue $\lambda^0_0$ must
vanish. On the other hand, considering that
\begin{eqnarray}
\phi_0^0(x)\propto \exp\left[-V(x)/2k_{\rm B}T\right],
\end{eqnarray}
and the square of $\phi_0^0$ is just the equilibrium Boltzmann
distribution $p_{\rm eq}(x)$ under the potential $V(x)$, we
rewrite the first correction of $\lambda_0(f)$ as
\begin{eqnarray}
\label{averagerate}
\lambda^{(1)}_0(f)=\int p_{\rm eq}(x)k_{\rm off}(x,f)dx,\\
p_{\rm eq}(x)\propto \exp\left[-V(x)/k_{\rm B}T\right].
\label{equilibriumdistribution}
\end{eqnarray}
Substituting the above formulaes to Eq.~\ref{transform}, the
probability density function then is approximated to be
\begin{eqnarray} p(x,t)\approx
N_0\exp\left(-\frac{V}{2k_{\rm
B}T}\right)\exp[-\lambda_0(f)t]\phi_0(f).
\end{eqnarray}

The quantity measured in the constant force experiments usually is
the mean lifetime of the bond,
\begin{eqnarray}
\label{averagelifetime}
\langle\tau\rangle(f)=-\int_0^{\infty}t\frac{dQ}{dt}dt=\int_0^{\infty}Q(t)dt,
\end{eqnarray}
where the survival probability $Q(t)$ is related to the
probability density as
\begin{eqnarray}
\label{survivalprobability}
Q(t)&=& \int  p(x,t)dx \nonumber\\
&\approx&\exp\left[-t\left(\lambda_0^{(1)}(f)+\cdots\right)\right].
\end{eqnarray}
Therefore the decay of $Q(t)$ is almost a monoexponential in large
$D$ limits.

{\bf The dynamic force mode.} In addition to the constant force
mode, force could be dynamic, {\it e.g.}, force increasing with a
constant loading rate in the biomembrane force probe (BFP)
experiment~\cite{Evans2004}. This scenario should be more
complicated than that for the constant force case. We still use
Eq.~\ref{origindiffusionequation} to describe the bond
dissociations by a dynamic force but the constant force therein is
replaced by a time-dependent function $f_t$. The initial condition
is the same as before. We firstly convert the diffusion-reaction
equation into the Schr$\ddot o$dinger-like presentation again,
\begin{eqnarray}
\label{timedependentSchodingerlikeequation} \frac{\partial
\phi}{\partial t}&=&{\cal H}_{f_t}\phi \nonumber\\
&=& -\left({\cal H} + k_{\rm off}(x,f_t)\right)\phi.
\end{eqnarray}
Please note that now we face a time-dependent Sch$\ddot o$dinger
operator, ${\cal H}_{f_t}$.

We know that solving a Sch$\ddot o$dinger equation with a
time-dependent Hamiltonian is difficult. The most simplest
situation is to use an adiabatic approximation analogous to what
is done in quantum mechanics. Hence we immediately
have~\cite{Messiah}
\begin{eqnarray}
\phi(x,t)\approx \exp\left[-\int_0^t \left(\lambda_0(f_{t'})+
B(t')\right)dt' \right]\phi_0(f_t),
\end{eqnarray}
where the ``Berry phase",
\begin{eqnarray}
B(t)=\int\phi_0(f_t) \frac{\partial}{\partial t}\phi_0(f_t)dx,
\end{eqnarray}
and $\phi_0(f_t)$ is the first eigenfunction of ${\cal H}_{f_t}$.
We are not very sure whether the assumption is reasonable or not
in practice. But one of tests is to see the agreement between the
calculation and data. We apply the perturbation approach again to
obtain the eigenvalues and eigenfunctions of the time-dependent
operator. Hence, we have $\phi_0(f_t)$ and $\lambda_0(f_t)$ by
only replacing the Bell expression in
Eqs.~\ref{eigenfunctionexpansion} and \ref{eigenvalueexpansion}
with $k_{\rm off}(x,f_t)$. The Berry phase then is approximated to
\begin{eqnarray}
\label{Berryphase}
B(f_t)&\approx&\sum_{m\neq0}\left(\frac{1}{\lambda^0_m}\right)^2\int\phi^0_0(x)k_{\rm
off}(x,f_t)\phi^0_m(x)
dx\times\nonumber\\
&&\int \phi^0_0(x)\frac{dk_{\rm off}}{dt}\phi^0_m(x) dx.
\end{eqnarray}
Finally, the survival probability for the dynamic force is given
by
\begin{eqnarray}
&&Q(t)\approx\exp\left[-\int_0^t\left(\lambda^{(1)}_0(f_{t'})+
B(f_{t'})+\cdots\right)dt'\right].\nonumber\\
\end{eqnarray}

Different from the constant force mode, the data of the dynamic
force experiments are typically presented in terms of the force
histogram, which corresponds to the probability density of the
dissociation forces $p(f)$
\begin{eqnarray}
p(f)=-\frac{dQ}{dt}\left/\frac{df}{dt}\right.
\end{eqnarray}
Particularly, when force is a linear function of time $f=f_0+rt$,
where $r$ is loading rate, and zero or nonzero of $f_0$
respectively corresponds to the steady- or jump-ramp force
mode~\cite{Evans2004}, we have
\begin{eqnarray}
\label{ruputureforcedistribution} &&P(f,f_0)\approx\frac{1}{r}
\left[\lambda^{(1)}_0(f)+B(f)+\cdots \right]\times\\
&&\exp\left[-\frac{1}{r}\int_{f_0}^f\left(\lambda^{(1)}_0(f')+
B(f')+\cdots\right)df'\right].\nonumber
\end{eqnarray}

\subsection{Slow diffusion}
At lower temperatures (or higher viscosities) the conformational
coordinate is, or the diffusion coefficient $D\to 0$. In this
limit, the solution of Eq.~\ref{origindiffusionequation} for the
constant force mode is simply given by
\begin{eqnarray}
\label{zerodiffusiondensity} p(x,t)=p_{\rm eq}(x)\exp\left[-k_{\rm
off}(x,f)t\right].
\end{eqnarray}
Compared to the density at larger diffusivity, the probability
density in this limit closely depends on the initial distribution.
Consequently, the survival probability $Q(t)$ as an integral of
$p(x,t)$ is a multiexponential decay which is remarkably distinct
from the monoexponential decay at larger $D$ limitation
(Eq.~\ref{survivalprobability}). The mean lifetime of bond is
given as
\begin{eqnarray}
\label{nondifussionmeanlifetime} \langle\tau\rangle(f)=\int dx
p_{\rm eq}(x)k^{-1}_{\rm off}(x,f).
\end{eqnarray}
For the dynamic force mode, the probability density is very
simple,
\begin{eqnarray}
\label{zerodiffusiondensity} p(x,t)=p_{\rm
eq}(x)\exp\left[-\int_0^tk_{\rm off}(x,f_{t'})dt'\right],
\end{eqnarray}
and the probability density of the dissociation force for the
ramping force at the slower diffusion is
\begin{eqnarray}
\label{nondifusionforcedissociationdistribution} P(f,f_0)&=&\int
dx p_{\rm eq}(x)\frac{1}{r}k_{\rm
off}(x,f)\times\nonumber\\
&&\exp\left[-\frac{1}{r}\int_{f_0}^fk_{\rm off}(x,f')df'\right].
\end{eqnarray}

When the diffusion coefficients are between the two limiting
cases, in general there are not analytical solutions to
Eq.~\ref{origindiffusionequation}. We have to employ numerical
approach (SSDP ver. 2.6)~\cite{Krissinel}.

\section{Two typical examples }
In the present work, we only consider the bound diffusions in a
harmonic potential
\begin{eqnarray}
\label{Harmonicpotential} V(x)=\frac{k_x}{2} x^2,
\end{eqnarray}
where $k_x$ is the spring constant. $\cal H$ then reduces to a
harmonic oscillator operator with an effective potential
\begin{eqnarray}
U(x)=\frac{Dk_x}{2k_{\rm B}T}\left(\frac{k_xx^2}{2k_{\rm
B}T}-1\right).
\end{eqnarray}
Its eigenvalues and eigenfunctions are
\begin{eqnarray}
\label{unpertubatedeigenvaluesharmonic} \lambda^0_n&=&nDk_x/k_{\rm
B}T
\end{eqnarray}
and
\begin{eqnarray}
\phi^0_n(y)&=&2^{-n/2}\pi^{-1/4}(n!)^{-1/2}e^{-z^2/2}H_n(y),
\end{eqnarray}
respectively, where $y=(k_x/2k_{\rm B}T)^{1/2}x$, and $H_n(y)$ is
the Hermite polynormials~\cite{Messiah}. In principle, we can
construct various functions of the barrier heights and projection
distances with respect to the conformational coordinate $x$. In
the present work, we only study the two examples, the linear
functions and the piecewise functions with two segments. The
former should be the simplest functional dependence on the
coordinate. While the later is complex enough to compare with the
real data.

\subsection{Rapid diffusion limit: $D\to \infty$}
{\bf Linear $\Delta G^\ddag$ and $\xi^\ddag$}. In this limit, the
higher order corrections $\lambda_0^{(n)}, (n\ge 2)$ are
negligible, and the survival probability of bond,
Eq.~\ref{survivalprobability}, is monoexponential decay.
Consequently, the mean lifetime of a bond is simply
$\langle\tau\rangle(f)=1/\lambda^{(1)}_0(f)$. We start from the
case that $\Delta G^\ddag$ and $\xi^\ddag$ are linear functions of
the conformational coordinate,
\begin{eqnarray}
\label{linearbarrierdistance}
\Delta G^\ddag(x)&=&\Delta G^\ddag_0+k_g x,\nonumber\\
\xi^\ddag(x)&=&\xi^\ddag_0+k_\xi x.
\end{eqnarray}
Substituting the above equations and the harmonic potential into
Eq.~\ref{averagerate}, we have
\begin{eqnarray}
\label{ratelinearbarrierdistance} \lambda_0(f) &=& k_0 \exp\left[
-\beta \Delta
G^\ddag_0+\frac{\beta}{2}\frac{k_g^2}{k_x}-\frac{\beta
k_x}{2k_\xi^2} \left(
\frac{k_gk_\xi}{k_x}-\xi^\ddag_0\right)^2\right]\nonumber \\
&&\times \exp\left[\frac{\beta
k_\xi^2}{2k_x}\left(f-\frac{k_gk_\xi/k_x-\xi^\ddag_0}{k_\xi^2/k_x}\right)^2\right],
\label{avgrate}
\end{eqnarray}
where $\beta=1/k_{\rm B}T$. We rewrite the above equation into a
more compact form,
\begin{eqnarray}
\label{GSRMlikerate} \lambda_0(f) &=& k_0 \exp\left[ -\beta \Delta
G^{\ddag}_0+\frac{\beta^2}{2}K_{g}-\frac{(\xi^{\ddag}_0-\beta
K_{g\xi})^2}{2K_{x}}\right]\nonumber \\
&&\times \exp\left[\frac{\beta^2K_\xi}{2}\left(f-\frac{\beta
K_{g\xi}-\xi^{\ddag}_0}{\beta K_\xi}\right)^2\right],
\label{avgrate}
\end{eqnarray}
by defining new variables as following,
\begin{eqnarray}
K_\xi=\frac{k_\xi^2}{\beta k_x},\text{ }K_g=\frac{ k_g^2}{\beta
k_x},\text{ }K_{g\xi}=\frac{k_\xi k_g}{\beta k_x}.
\end{eqnarray}
Eq.~\ref{GSRMlikerate} has the same expression with the average
dissociation rate formally obtained in the GSRM (Eq. 6 in
Ref.~\cite{liuf1} and $x^\ddag_0$ instead of $\xi^\ddag_0$ used
therein). Consequently, $K_g$, $K_\xi$ and $K_{g\xi}$ could be
viewed as the variances and covariance of and between the two
stochastic variables $\Delta G^\ddag$ and $\xi^\ddag$ at the same
time point. Interestingly, it also means that we cannot get the
absolute value of the parameters $k_x$, $k_g$ and $k_\xi$ by
observing the force dependence of the dissociation rates. Although
the above equation is not new, there are still several points
deserving to be presented. Firstly, the derivation of
Eq.~\ref{GSRMlikerate} does not require the forced bond
dissociation process stationary as the GSRM assumed. Then the
correlation coefficient of the two stochastic variables is only 1
or $-1$ according to whether $k_g$ and $k_\xi$ have the same signs
or not, whereas in the GSRM the coefficient is arbitrary. It is
expected because both $\Delta G^\ddag$ and $\xi^\ddag$ here are
the linear functions of the same coordinate. Finally,
$\xi^\ddag_0$ could be minus. We are not ready to discuss the
general behaviors of Eq.~\ref{GSRMlikerate} in detail because it
has been performed in the GSRM. We only emphasize that the rate
indeed exhibits the catch-slip bond transitions, when the
following conditions hold: both the distance and the barrier
height are stochastic variables, and $\xi^\ddag_{\rm c}=\beta
K_{g\xi}-\xi^{\ddag}_0>0$. Because the force is always positive,
the rate then will firstly decreases with the increasing of the
force, and then increases when the force is beyond a threshold,
$f_{\rm c}=\xi^\ddag_{\rm c}/\beta K_\xi$. We define the new
parameters $\xi^\ddag_{\rm c}$ and $f_{\rm c}$ since they have the
same dimensions of distance and force.
\begin{figure}[htpb]
\begin{center}
\includegraphics[width=0.7\columnwidth]{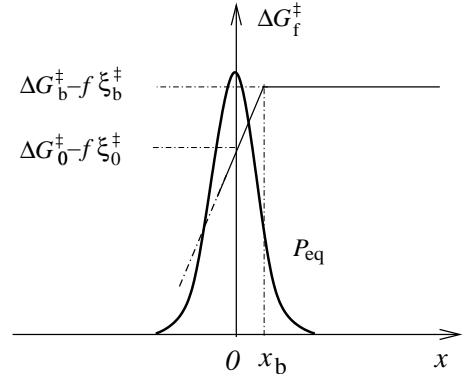}
\caption{Schematic diagram of the piecewise functions $\Delta
G^\ddag_{\rm f}=\Delta G^\ddag(x)-f\xi^\ddag(x)$ with respect to
the coordinate $x$ at a given force $f$. The symmetric curve
represents the thermal equilibrium distribution of the
conformational coordinate under the harmonic potential $V(x)$.}
\label{figure1}
\end{center}
\end{figure}

Following the GSRM, we combine the six parameters in
Eq.~\ref{GSRMlikerate} into only three: the threshold $f_{\rm c}$,
the variance $K_\xi$, and a prefactor $N$ which is defined as
\begin{eqnarray}
N=k_0\exp{\left[ -\beta \Delta
G^{\ddag}_0+\frac{\beta^2}{2}K_{g}-\frac{\beta^{2}}{2}K_\xi
(f-f_c)^2\right]}
\end{eqnarray}
Then the mean lifetime has a Gaussian-like function with respect
to force,
\begin{eqnarray}
\label{Gaussianlikelifetime} \langle \tau\rangle(f)
&=&N^{-1}\exp\left[-\frac{\beta^{2}}{2}K_\xi (f-f_c)^2\right].
\end{eqnarray}
We respectively fit the data of the forced dissociations of the
single PSGL-1$-$P-~\cite{Marshall} and
L-selectins~\cite{Sarangapani} bonds. The fitting results and the
parameters will be showed in the next section. Because the
Gaussian-like expression is symmetric relative to the threshold
$f_{\rm c}$, and it is clearly inconsistent with the experimental
observations, in the following section, we study a slightly
complicated case.

\begin{figure}[htpb]
\begin{center}
\includegraphics[width=1\columnwidth]{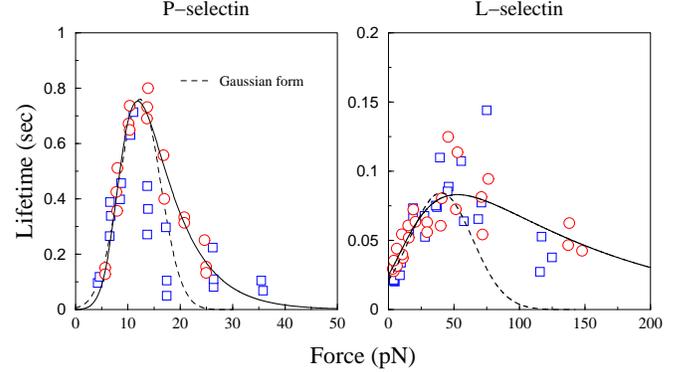}
\caption{The mean lifetime vs. force for the bonds of dimeric
P-selectin and L-selectin with the monomeric sPSGL-1 (the
square)~\cite{Marshall} and the dimeric PSGL-1 (the circle). The
reader is reminded that the data for the dimeric
PSGL-1$-$P-selectin bonds has been rescaled to single bond
case~\cite{Pereverzev}. The solid lines are the results of
Eq.~\ref{ratebendlandscape}, where the parameters are as follows:
for P(L)-selectin case, $\kappa_0\approx0.09$ (4.60) s$^{-1}$,
$K_g\approx18.5$ (3.39) ${k_{\rm B}T}^2$, $K_\xi\approx2.0$ (0.06)
nm$^{2}$, and $f_{\rm c}=11.0$ (27.7) pN.  While the dashed lines
are from the Gaussian-like Eq.~\ref{GSRMlikerate}, where the
parameters are as follows: for P(L)-selectin case, $N\approx0.76$
(0.085) s$^{-1}$, $K_\xi\approx1.0$ (0.03) nm$^2$, and $f_{\rm
c}\approx 12.3$ (40.0) pN. Compared to the mean lifetime vs. force
in the FMDD model, the present curve decays slowly after the
transition. } \label{figure2}
\end{center}
\end{figure}

{\bf Piecewise $\Delta G^\ddag$ and $\xi^\ddag$ functions.}
Compared to the former, the current case should be more useful
because it describes the experiments better. We assume that both
the barrier and distance are piecewise function with two segments:
for $x\le x_b$, they follow Eq.~\ref{linearbarrierdistance};
otherwise,
\begin{eqnarray}
\Delta G^\ddag(x)&=&\Delta G^\ddag_b=\Delta G^\ddag(x_b),\nonumber\\
\xi^\ddag(x)&=&\xi^\ddag_{b}=\xi^\ddag(x_b).
\end{eqnarray}
Fig.~\ref{figure1} is the schematic diagram of the function
$\Delta G^\ddag(x)-f\xi^\ddag(x)$ at a given force. Substituting
the functions into Eq.~\ref{averagerate}, we have
\begin{eqnarray}
\label{ratebendlandscape} \lambda_0(f)&=&\frac{k_0}{2}
\exp\left[-\beta \Delta G_0^\ddag+\frac{\beta
k_g^2}{2k_x}-\frac{\beta k_x}{2k_\xi^2}\left(\frac{k_\xi
k_g}{k_x}-\xi_0^\ddag\right)^2\right] \nonumber \\
&& \times\exp\left[\frac{\beta
k_\xi^2}{2k_x}\left(f-\frac{k_gk_\xi/k_x-\xi^\ddag_0}{k_\xi^2/k_x}\right)^2\right]
\times \\
&&{\rm erfc}\left[-\sqrt{\frac{\beta
k_x}{2}}\left(x_b+\frac{k_g}{k_x}-\frac{k_\xi}{k_x}f\right)\right]
+\nonumber\\
&&\frac{k_0}{2}\exp\left[-\beta \Delta G^\ddag_b+\beta
f\xi^\ddag_b\right] \times {\rm erfc}\left[\sqrt{\frac{\beta
k_x}{2}}x_b\right]\ \nonumber
\end{eqnarray}
where the complementary error function is
\begin{eqnarray}
{\rm erfc}(x)=\frac{2}{\sqrt{\pi}}\int_x^\infty e^{-x^2}dx.
\end{eqnarray}
Although the above rate expression seems to be more complex than
that in the case of the linear functions, it is really a simple
weighted sum of Eq.~\ref{ratelinearbarrierdistance} and the
standard Bell expression, Eq.~\ref{Bellmodel}. This function would
yield the major characteristics of the catch-slip bond transitions
if we assumed that $k_\xi>0$, $\xi^\ddag_b>0$, and $k_xx_b^2\gg0$:
(i) if force is smaller, Eq.~\ref{ratebendlandscape} reduces to
Eq.~\ref{ratelinearbarrierdistance} because the asymptotic
behaviors of the error function are
\begin{eqnarray}
{\rm erfc}(-\infty)=2, \qquad {\rm erfc}(+\infty)=0,
\end{eqnarray}
respectively; (ii) if force is larger enough, according the
asymptotic expansion of the error function for large positive $x$,
\begin{eqnarray}
{\rm erfc}(x)\approx \frac{e^{-x^2}}{\sqrt
\pi}\left(\frac{1}{x}-\frac{1}{2x^3}+\cdots\right),
\end{eqnarray}
it is easy to prove that Eq.~\ref{ratebendlandscape} tends to the
standard Bell expression with a positive projection distance. This
behavior has been observed in experiment~\cite{Alon} early. To fit
the data, we rewrite Eq.~\ref{ratebendlandscape} into
\begin{eqnarray}
\label{Sreactionratebendlandscape} \lambda_0(f) &=&
\kappa_0\times \left\{\right.\nonumber\\
&&\exp\left[ \frac{\beta^2K_g}{2}-\frac{\beta^2K_\xi}{2} f_c^2\right]
\times\exp\left[\frac{\beta^2K_\xi}{2}(f-f_c)^2\right]\nonumber \\
&&\times{\rm erfc}\left[-X_b-\sqrt{\frac{\beta^2K_g}{2}}+
\sqrt{\frac{\beta^2K_\xi}{2}}f \right]
+\nonumber\\
&&\exp\left[-2
X_b\sqrt{\frac{\beta^2K_g}{2}}+f\left(\beta^2K_{g\xi}^2
-\right.\right.\nonumber\\
&&\left.\left.\left. \beta^2K_\xi
f_c+2X_b\sqrt{\frac{\beta^2K_\xi}{2} }\right)\right]\times {\rm
erfc}[X_b] \right\}, \nonumber
\end{eqnarray}
by introducing additional definitions,
\begin{eqnarray}
X_b&=&x_b\sqrt{\frac{\beta k_x}{2}},\text{
}\kappa_0=\frac{k_0}{2}\exp\left(-\beta \Delta G_0^\ddag\right),
\end{eqnarray}
where we have replaced $\Delta G^\ddag_b$ and $\xi^\ddag_b$ by the
linear functions. We see that we cannot determine parameters
$k_x$, $k_g$ and $k_\xi$ by fitting the forced dissociation
experiments. Totally there are five parameters presenting in the
current model, $K_\xi,\text{ }K_g,\text{ }X_b,\text{ }\kappa_0$,
and $f_c$. In practice, we further reduce the model to the
simplest case by assuming $X_b=0$. The fitting curves for the
forced dissociations of PSGL-1$-$P-selectin and -L-selectin
complex are showed in Fig.~\ref{figure2}. We see that the
agreement between the model and the data is quite good.

\begin{figure}[htpb]
\begin{center}
\includegraphics[width=0.9\columnwidth]{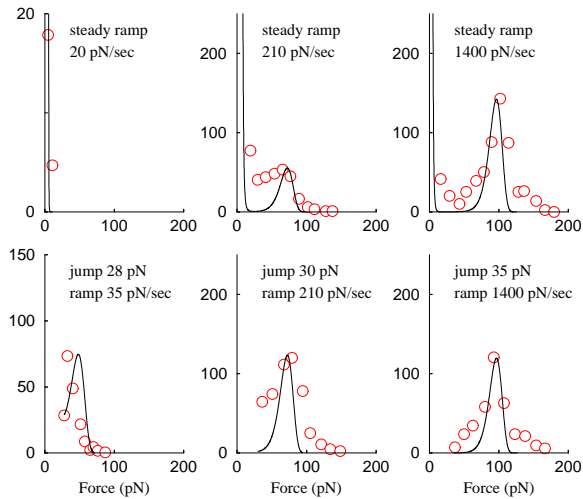}
\caption{The probability density of the dissociation forces
$P_{\rm d}(f,f_0)$ under the different loading rates predicted by
our theory (solid curves) for the PSGL-1$-$P-selectin complex
(symbols)~\cite{Evans2004}. The reader is reminded that each curve
requires an adjustable normalization parameter to fit each
histogram. }\label{figure3}
\end{center}
\end{figure}

More challenging experiments to our theory are the force steady-
and jump-ramp modes~\cite{Evans2004}. In $D\to\infty$ limit,
Eq.~\ref{ruputureforcedistribution} reduces to
\begin{eqnarray}
\label{ruptureforceprobability}
P(f,f_0)=\frac{\lambda_0(f)}{r}\exp \left
[-\frac{1}{r}\int_{f_0}^f\lambda_0(f')df' \right].
\label{approxruputureforcedistribution}
\end{eqnarray}
According Eq.~\ref{ruptureforceprobability}, the probability
density of dissociation force is uniquely determined by the rate
$\lambda_0(f)$. Because both the present work and FMDD model
describe the same group of experiments, here we are not ready to
discuss the general characteristics of $P(f,f_0)$; they have been
done in FMDD~\cite{liuf2}. When we test
Eq.~\ref{ruptureforceprobability} with the parameters obtained
from the constant force data, it apparently deviates from the
dynamic force data~\cite{Evans2004}. It must emphasize that the
experiments of the constant and dynamic force modes were performed
by the same experimental group and the same \emph{single} bond
sPSGL-1$-$P-selectin. Similar problem has been met in our previous
FMDD model and two-pathway and one well model~\cite{Pereverzev}.
We~\cite{liuf2} argued and suggested that the experiment of
dynamic force modes really measured the histogram of the
dissociation force of the \emph{dimeric} bond PSGL-1$-$P-selectin
complex instead of single bond case: the two independent bonds
share the same force and fail randomly. We still make use of this
assumption in the present work. It is easy to relate the
probability density of the dissociation force for the dimeric
bonds, $P_{\rm d}(f,f_0)$, to the single one by
\begin{eqnarray}
\label{equivalentbond} P_{\rm d}(f,f_0)=P\left(f/2,f_0/2\right)^2.
\end{eqnarray}
Fig.~\ref{figure3} is the finally result.
We see the theoretical prediction is satisfactory. Of course,
further experiments are needed to prove the correctness of this
dimeric bonds assumption.

\subsection{Intermediate diffusivity: numerical approach }
Even if the Bell rate model with dynamic disorder can well
describe the experimental data in quantitative and qualitative
ways, it does not mean that our model is real; Our theory might
only work in a specific experimental condition, $e.g.$, in the
larger $D$ limit. We should test the model in a broad range of
experimental conditions. Although until now there are not new
experiments about the catch-slip bond transitions, we
theoretically study the general behaviors of
Eq.~\ref{origindiffusionequation}. In this and next sections, we
consider the Bell rate model in the presence of dynamic disorder
at intermediate and zero $D$, respectively. In principle, the
existing experiments would be easily modified to test our
prediction, such as by decreasing experimental temperature,
increasing viscosity of solvent, and etc. Different from the
limiting cases, we solve Eq.~\ref{origindiffusionequation} for
intermediate diffusion coefficients $D$ via a numerical computing
software (SSDP ver.2.6)~\cite{Krissinel}. Here we focus on the
piecewise barrier and distance case because. We give the
reasonable parameters, $\kappa=$0.09 s$^{-1}$, $k_x=$1.0 pN
nm$^{-1}$, $k_g=$8.61 pN, $k_\xi=$0.70, $\xi^\ddag_0=0.64$ nm, and
$x_b=0$ nm, which can recover the parameters applied to fit the
data of PSGL-1$-$P-selectin bond. The initial distribution is set
equal to the equilibrium distribution with the harmonic potential
Eq.~\ref{Harmonicpotential}.

\begin{figure}[htpb]
\begin{center}
\includegraphics[width=1\columnwidth]{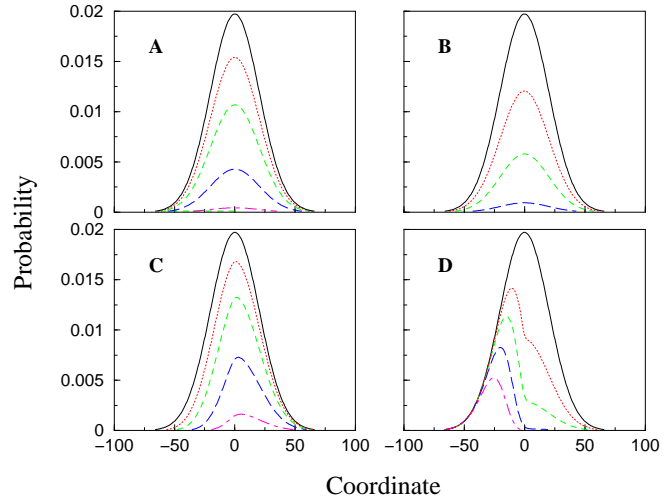}
\caption{Probability density $p(x,t)$ at five times for two forces
and diffusion coefficients: $D=10^{-10}$ and $10^{-15}$ cm$^2/$s
for the first and second row, and $f=10$ and 25 pN for the first
and second column, respectively. The times are: 0, 175, 937, 1090,
and 2700 ms. The distribution at 2700 ms in the panel B are too
small to be showed. } \label{figure4}
\end{center}
\end{figure}

\begin{figure}[htpb]
\begin{center}
\includegraphics[width=1\columnwidth]{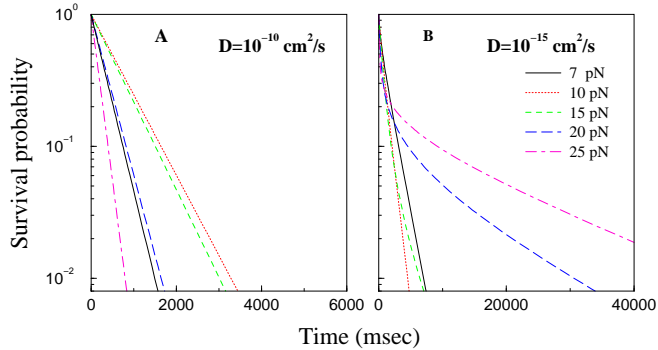}
\caption{Survival probabilities vs. time at five forces for the
two diffusion coefficients. } \label{figure5}
\end{center}
\end{figure}

We calculate $p(x,t)$ for two diffusion coefficients $D$ and two
forces; see Fig.~\ref{figure4}.
When the diffusion coefficient is small, the diffusion along $x$
is practically frozen. Hence $p(x,t)$ will decay more on the side
that k(x) is larger. Because both $\Delta G^\ddag$ and $\xi^\ddag$
are piecewise functions with two segments, we easily estimate the
force, $f_{\rm e}=k_g/k_\xi$, at which all decay rates are
independent of the coordinate $x$. Because f=10 pN and 25 pN are
respectively smaller and larger than $f_{\rm e}=12.3$ pN given the
current parameters, for the smaller force, the $p(x,t)$ decays
more on the left side, while for the larger force, $p(x,t)$ decays
more on the right side. Fig.~\ref{figure4}C and D indeed indicate
this prediction. In contrast, if the diffusion coefficient is
large, the dissociation is slow compared to process of the
conformational diffusion. The thermal equilibrium distribution of
the conformations should be maintained during the courses of the
dissociation ``reaction". It is cleanly seen in
Fig.~\ref{figure4}A and B. Because the single-molecule forced
dissociation experiments usually measure the bond survival
probability, we calculate $Q(t)$ for the same two diffusion
coefficients and five forces; see Fig.~\ref{figure5}.
For the larger diffusion coefficient (Fig~\ref{figure5}C), $Q(t)$s
are almost monoexponential decays with time. We see that the decay
first slows down with force increasing initially, and then speeds
up when the force is beyond a threshold. These properties have
been discussed in the rapid diffusion limit. More interesting case
is from the smaller diffusion coefficients, {\it e.g.},
$D=10^{-15}$ cm$^2$s$^{-1}$ in Fig.~\ref{figure5}D, where the main
characteristics are completely different from the larger $D$ case.
First, the survival probabilities exhibit highly stretched
multiexponential decays. The trend becomes more apparent at larger
forces. Second, the decay first speeds up and then slows down with
force increasing. This abnormal behavior surprises us. It possibly
means that the catch-slip bond transition in rapid diffusion limit
converts into slip-catch transition when the diffusion is very
slow along the conformational coordinate. In order to understand
this behavior better, we consider the nondiffusion limit in which
Eq.~\ref{origindiffusionequation} has analytic solution.

\subsection{nondiffusion limit: $D\to 0$}
\begin{figure}[htpb]
\begin{center}
\includegraphics[width=0.8\columnwidth]{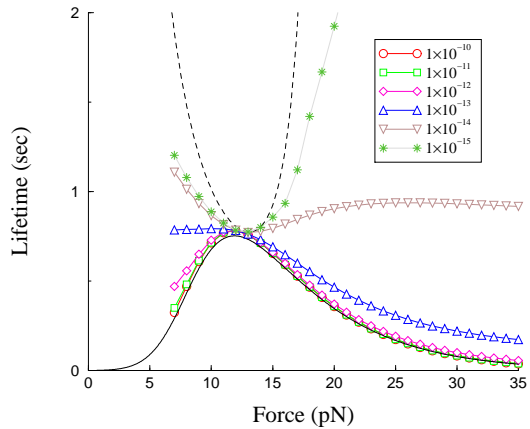}
\caption{Mean lifetimes of the PSGL-1$-$P-selectin bond for six
diffusion coefficients. The solid and dashed curves are
respectively calculated by Eqs.~\ref{ratebendlandscape}
and~\ref{ratebendlandscapenondiffusion} in $D\to\infty$ and
$D\to$0 limits.} \label{figure6}
\end{center}
\end{figure}

In this limit, according to Eq.~\ref{nondifussionmeanlifetime}, we
get the mean lifetimes for the linear barrier and projection
distance case (Eq.~\ref{linearbarrierdistance}),
\begin{eqnarray}
\langle\tau\rangle(f) &=& k_0^{-1} \exp\left[ \beta \Delta
G^\ddag_0+ \frac{\beta}{2}\frac{k_g^2}{k_x}-
\frac{\beta k_x}{2k^2_\xi}\left(\frac{k_gk_\xi}{k_x}+\xi^\ddag_0\right)^2\right] \nonumber\\
&&\times \exp\left[\frac{\beta
k_\xi^2}{2k_x}\left(f-\frac{k_gk_\xi/k_x+\xi^\ddag_0}{k_\xi^2/k_x}\right)^2\right],
\label{avgrate}
\end{eqnarray}
and for the piecewise barrier and projection distance case,
\begin{eqnarray}
\label{ratebendlandscapenondiffusion}
\langle\tau\rangle(f)&=&\frac{k_0^{-1}}{2}\exp\left[\beta \Delta
G_0^\ddag+\frac{\beta}{2}\frac{k_g^2}{k_x}-
\frac{\beta k_x}{2k^2_\xi}\left(\frac{k_gk_\xi}{k_x}+\xi^\ddag_0\right)^2\right]\nonumber \\
&&\times \exp\left[\frac{\beta
k_\xi^2}{2k_x}\left(f-\frac{k_gk_\xi/k_x+\xi^\ddag_0}{k_\xi^2/k_x}\right)^2\right]
\times\nonumber\\
&&{\rm erfc}\left [-\sqrt{\frac{\beta k_x}{2} }\left(x_b -
\frac{k_g}{k_x}+\frac{k_\xi}{k_x}f\right)\right ]
+\\
&&\frac{k_0^{-1}}{2}\exp\left[\beta \Delta G_0^\ddag -\beta
f\xi_b^\ddag \right]  {\rm erfc}\left[\sqrt{\frac{\beta
k_x}{2}}x_b\right] \nonumber,
\end{eqnarray}
respectively. We see that, when force is larger enough,
$k_{\xi}>0$ required in $D\to\infty$ case leads the mean lifetime
of the piecewise case in the nondiffusion limit to increase
rapidly in an exponential way of $f^2$. It proves our estimation.
In addition, for the case of the linear barrier and projection
distance, this catch behavior at larger forces even does not
depend on any parameters. Fig.~\ref{figure6} shows how the mean
lifetime of the PSGL-1$-$P-selectin bond changes with the decrease
of $D$,
where same parameters are used. Compared to the counterintuitive
catch-slip bond transitions, the slip-catch transitions presented
here is physically inconceivable.

\section{conclusions}
Stimulated by the discovery of the catch-slip bond transitions in
some biological adhesive bonds, several theoretical models have
been developed to account for the intriguing phenomenon. According
to the difference of the physical pictures, they can be divided
into two classes. The first class is based on discrete chemical
reaction schemes. They all assumed that there were two pathways
and two or one energy well~\cite{Evans2004,Barsegov,Pereverzev}.
The applied force alters the fractions of the two pathways. The
other class is based continuum diffusion-reaction models and
mainly proposed by us~\cite{liuf1,liuf2}. We cannot definitely
distinguish which theory or model is the most reasonable and more
close real situations, because their theoretical calculations all
agree with the existing experimental data. In our opinion, the
continuum models may be more attractive. In addition that the
chemical kinetic models are clearly oversimplified, the most
important cause is that the continuum models include an additional
parameter, the diffusion coefficient $D$~\cite{liuf2}. The
presence of the coefficient obliges the continuum
diffusion-reaction models to face a strict test: it must be
physically reasonable in the whole range of $D$. One of examples
is that the present Bell rate model with dynamic disorder may be
physically unacceptable because it predicts a slip-catch
transition when the coefficient tends to zero. We have pointed our
that the discrete chemical schemes may a mathematical
approximation of the continuum model~\cite{liuf2}. It is easy to
see that the two pathway and one energy well model with a minus
projection distance~\cite{Pereverzev} just corresponds the present
model in the rapid diffusion limit. Therefore it is very possible
that this chemical kinetic model also has a similar flaw. In
contrast, FMDD model passes this test~\cite{liuf2}. Of course,
future experiments will
finally decide which model is correct.\\
\\
This work was supported in part by the National Science Foundation
of China and the National Science Foundation under Grant No.
PHY99-07949.

 \end{document}